\documentclass[aip, jap, reprint, showkeys, showpacs]{revtex4-1}
\usepackage{graphicx}
\usepackage{amsmath}
\usepackage[latin1]{inputenc}
\usepackage[usenames,dvipsnames]{color}
\makeatletter
\DeclareRobustCommand*\textsubscript[1]{%
  \@textsubscript{\selectfont#1}}
\def\@textsubscript#1{%
  {\m@th\ensuremath{_{\mbox{\fontsize\sf@size\z@#1}}}}}
\makeatother
\begin{document}
\title{Formation and annealing of dislocation loops induced by nitrogen implantation of ZnO}
\author{Guillaume Perillat-Merceroz}
\email{guillaume.perillat-merceroz@cea.fr}
\affiliation{CEA, LETI, MINATEC Campus, 17 rue des Martyrs, 38054 Grenoble cedex 9, France}
\affiliation{CEA, INAC, SP2M, LEMMA, 17 rue des Martyrs, 38054 Grenoble cedex 9, France}
\author{Patrice Gergaud}
\affiliation{CEA, LETI, MINATEC Campus, 17 rue des Martyrs, 38054 Grenoble cedex 9, France}
\author{Pascal Marotel}
\affiliation{CEA, LETI, MINATEC Campus, 17 rue des Martyrs, 38054 Grenoble cedex 9, France}
\author{Stephane Brochen}
\affiliation{CEA, LETI, MINATEC Campus, 17 rue des Martyrs, 38054 Grenoble cedex 9, France}
\author{Pierre-Henri Jouneau}
\affiliation{CEA, INAC, SP2M, LEMMA, 17 rue des Martyrs, 38054 Grenoble cedex 9, France}
\author{Guy Feuillet}
\affiliation{CEA, LETI, MINATEC Campus, 17 rue des Martyrs, 38054 Grenoble cedex 9, France}
\date{\today}
\begin{abstract}
Although zinc oxide is a promising material for the fabrication of short wavelength optoelectronic devices, \textit{p}-type doping is a step that remains challenging for the realization of diodes. Out of equilibrium methods such as ion implantation are expected to dope ZnO successfully provided that the non-radiative defects introduced by implantation can be annealed out. In this study, ZnO substrates are implanted with nitrogen ions, and the extended defects induced by implantation are studied by transmission electron microscopy and X-ray diffraction (XRD), before and after annealing at $900\,^{\circ}\text{C}$. Before annealing, these defects are identified to be dislocation loops lying either in basal planes in high N concentration regions, or in prismatic planes in low N concentration regions, together with linear dislocations. An uniaxial deformation of 0.4\% along the c axis, caused by the predominant basal loops, is measured by XRD in the implanted layer. After annealing, prismatic loops disappear while the density of basal loops decreases and their diameter increases. Moreover, dislocation loops disappear completely from the sub-surface region. XRD measurements show a residual deformation of only 0.05\% in the implanted and annealed layer. The fact that basal loops are favoured against prismatic ones at high N concentration or high temperature is attributed to a lower stacking fault energy in these conditions. The coalescence of loops and their disappearance in the sub-surface region are ascribed to point defect diffusion. Finally, the electrical and optical properties of nitrogen-implanted ZnO are correlated with the observed structural features.
\end{abstract}
\pacs{61.72.U-, 61.72.Dd, 68.55.Ln}
\keywords{ZnO, ion implantation, annealing, semiconductor doping, transmission electron microscopy}
\maketitle
\section{Introduction}
Because of its direct band gap of 3.37~eV, zinc oxide has gained a renewed interest during the last decade due to a wide range of possible applications in short wavelength optoelectronics. ZnO is an interesting alternative to gallium nitride (which has approximatively the same band gap) for several reasons. The high binding energy of the exciton (59 meV for ZnO versus 30 meV for GaN) could provide better luminescence at room temperature. Contrary to GaN-based devices which contain relatively rare materials like Ga and In, ZnO-based devices contain Zn and Mg which are abundant and non-toxic elements. Finally, relatively cheap ZnO substrates are commercially available contrary to GaN ones. This allows homoepitaxial ZnO layers to be grown with less structural defects than GaN heteroepitaxial layers. Important progresses have been made in the growth of ZnO and its alloys under the form of bulk crystals, thin layers, and nanowires.\cite{ozgur_jap_2005} But one of the remaining issues for the fabrication of ZnO-based devices is obtaining stable and reproducible \textit{p}-type doping.\cite{janotti_rpp_2009} Group V elements such as N, P, As, or Sb substituting for O atoms are envisaged for \textit{p}-type doping. Among them N is theoretically the best candidate because of its higher electronegativity and its ionic radius similar to that of O. However, from recent \textit{ab-initio} calculations, the ionization energy could be 0.33~eV,\cite{duan_prb_2009} or 1.3~eV,\cite{lyons_apl_2009} indicating that \textit{p}-type doping with N may be difficult at thermodynamic equilibrium. Nevertheless, ZnO-based LEDs grown by molecular beam epitaxy at a temperature around $900\,^{\circ}\text{C}$ (a relatively low temperature compared to the melting point of $1975\,^{\circ}\text{C}$) have been demonstrated using nitrogen for \textit{p}-type doping.\cite{tsukazaki_nm_2005,nakahara_apl_2010} Ion implantation is also promising for \textit{p}-type doping of ZnO with nitrogen: it is a far from equilibrium method as a high supersaturation of point defects is created at room temperature, and N concentration can be beyond its equilibrium solubility. Moreover this method offers a precise doping through the control of dose, area localization, and dopant depth. For these reasons N\textsuperscript{+} implantation of ZnO (called hereafter N implantation) is actively studied.\cite{adekore_jap_2007,kennedy_jap_2010,reuss_jap_2004,wang_apl_2006,gu_apl_2008} However implantation of energetic ions introduces defects in the crystal lattice which degrade the electrical\cite{wang_apl_2006} and optical\cite{walsh_sse_1977} properties. Annealing is needed to remove these defects and to activate dopant atoms by moving them to the right lattice sites, keeping in mind that annealing at too high temperatures may on the contrary deactivate dopants by approaching equilibrium or by forming molecular nitrogen.\cite{fons_prl_2006} Although N-implanted and annealed ZnO remains \textit{n}-type in general,\cite{reuss_jap_2004,borseth_prb_2008,chen_jpcm_2004} donor-acceptor pair emission has been observed by photoluminescence (PL).\cite{meyer_pssb_2004,kennedy_jap_2010,reuss_jap_2004,marotel_pc_2010} Electroluminescence has been sometimes observed, but it is not clear if this was due to the formation of an actual p-n junction,\cite{adekore_jap_2007,gu_apl_2008} or to a metal-insulator-semiconductor junction.\cite{wang_apl_2006} These observations call for a thorough study of implantation defects and of their evolution upon annealing, in order to correlate the optical and electrical properties with the structural ones.
The defects induced in ZnO by irradiation vary according to the type of irradiation (ion or electron), and according to the implanted element. Voids are produced by As implantation,\cite{coleman_apl_2005} and clusters by Co or Mn implantation.\cite{zhou_prb_2008} Although ion implantation in ZnO does not usually introduce any amorphous layer because of the large ionicity of its chemical bounds,\cite{trachenko_prb_2005} Si implantation can cause ZnO amorphization.\cite{kucheyev_prb_2003} Ion bombardment thinning of heavily doped ZnO with low voltage Ar\textsuperscript{+} results in the formation of interstitial dislocation loops lying in basal planes with a Burgers vector $\mathbf{b} = \mathbf{c}/2 + \mathbf{p}$.\cite{couderc_mmm_1995} Electron irradiation of a ZnO thin foil creates interstitial dislocation loops lying in basal planes with $\mathbf{b} = \mathbf{c}/2$, and lying in prismatic planes with $\mathbf{b} = \mathbf{a}$.\cite{yoshiie_pma_1980} It seems that dislocation loops are also formed during Au implantation, but the authors note that a more detailed transmission electron microscopy (TEM) study would be needed.\cite{kucheyev_prb_2003} Defects induced by N implantation of ZnO, and sometimes their evolution with annealing, have also been studied. A Rutherford backscattering spectrometry study indicates that the absolute concentration of defects is much lower in ZnO than in GaN.\cite{lorenz_apl_2005} From X-ray diffraction (XRD), it is found that the crystalline quality is recovered after an annealing at $800\,^{\circ}\text{C}$.\cite{adekore_jap_2007} Positron annihilation spectroscopy studies reveal the formation of vacancy clusters, and explain their behaviour upon annealing.\cite{chen_jpcm_2004,chen_apl_2005, borseth_prb_2008,dong_prb_2010} Nevertheless, the crystallographic nature of the extended defects induced by N implantation of ZnO has never been determined, and the annealing effect has not been observed directly by TEM.
In this study, N implantations were carried out in order to study the formation of structural defects and their recovery upon annealing. TEM and XRD were used to study an as-implanted sample, and a sample after implantation and annealing. Extended defects are identified to be dislocation loops lying in basal or prismatic planes, together with linear dislocations. The evolution of the defects with annealing was observed. Induced deformations in the implanted layer were measured by XRD and correlated to the observed microstructure. Considerations on dislocation energy and on local N concentration give some elements to discuss the different habit planes of the loops and the annealing effects. Optical and electrical characterizations obtained on the same samples are correlated with these structural observations. N being a potential \textit{p}-type dopant for ZnO, the knowledge of the type of defects and their evolution with annealing is indeed of great importance in order to optimize structural, electrical, and optical properties of N-implanted ZnO.
\section{Experimental details}
Two hydrothermally grown single-crystals of ZnO provided by Crystec were studied: one is  an as-implanted substrate, and the other one is an implanted and annealed substrate. N\textsuperscript{+} implantations were carried out at $0\,^{\circ}\text{C}$ on the oxygen face (\textit{i.e.} the $(000\overline{1})$ crystal face) with a NV-8200P Axcelis implanter. Samples were tilted 7$^\circ$ relative to the ion beam direction to minimize channeling. The beam flux was $1.15 \times 10^{12}$~cm\textsuperscript{-2}s\textsuperscript{-1}. Three successive implantations at different energies and doses were carried out on each of the two samples in order to obtain a flat nitrogen profile. Accelerating voltages of 50, 120, and 200~kV with respective doses of $4\times10^{14}$, $8\times10^{14}$, and $10^{15}$~cm\textsuperscript{-2} were used. The second sample was annealed at $900\,^{\circ}\text{C}$ during 15 minutes under O$_2$ at atmospheric pressure. These annealing conditions were chosen because they lead to the maximum intensity for the donor-acceptor pair emission in PL spectra.\cite{marotel_pc_2010}
TEM sample preparation was carried out without ion milling in order to avoid artefacts, since Ar\textsuperscript{+} milling of ZnO at 5~kV is known to introduce defects.\cite{couderc_mmm_1995} Although they can be minimized by a final milling at low voltage,\cite{vennegues_jap_2008} the wedge polishing technique was preferred.\cite{voyles_um_2003} TEM images were taken on a Jeol 4000 EX microscope operated at 400~kV. Scanning TEM images were taken on a FEI Titan microscope fitted with a probe $C_{\text{S}}$-corrector and operated at 300~kV. XRD experiments were carried out on a Panalytical Xpert MRD diffractometer implementing triple axis measurements with the use of a fourfold Ge(220) monochromator, a threefold Ge(220) analyser, and Cu $K\alpha$ radiation (1.540598 Å).

The following crystallographic notations are used in this paper for hexagonal ZnO: $ \mathbf{c} =\langle0001\rangle$, $\mathbf{a} = \frac{1}{3} \langle11\overline{2}0\rangle$, $\mathbf{p} = \frac{1}{3} \langle01\overline{1}0\rangle$. $\{0001\}$ planes are called basal planes. $\{11\overline{2}0\}$ and $\{1\overline{1}00\}$ planes are called prismatic planes. Dislocation loops lying in basal or prismatic planes are called basal or prismatic loops respectively.  In the wurtzite structure, two types of basal loops can be found: $\mathbf{b}=\mathbf{c}/2$ dislocation loops which enclose extrinsic stacking faults with a stacking sequence ABABACBABA, and $\mathbf{b}=\mathbf{c}/2 + \mathbf{p}$ dislocation loops which enclose type one intrinsic stacking faults with a stacking sequence ABABABACACA.\cite{couderc_mmm_1995,hirth_1992}
\section{Results}
\subsection{N concentration profiles}
\begin{figure}
\includegraphics[width=8cm]{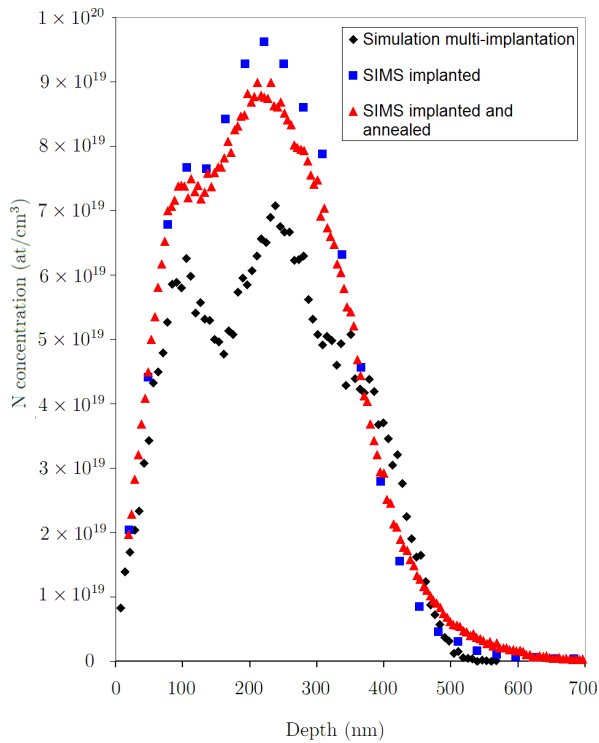}
\caption{\label{N_SIMS_SRIM}N concentration profiles simulated with the SRIM software,\cite{ziegler_nimpr_2004} and measured by SIMS on the as-implanted sample and on the implanted and annealed sample.}
\end{figure}
Fig.~\ref{N_SIMS_SRIM} shows the implantation profiles measured by secondary ion mass spectrometry (SIMS) before and after annealing, together with the simulated profile obtained with the {SRIM} software.\cite{ziegler_nimpr_2004} SIMS measurements and the simulation are in good agreement.The annealed and the as-implanted sample present nearly the same concentration profile. Thus N diffusion from the implanted zone in ZnO or to the surface appears to be negligible at $900\,^{\circ}\text{C}$ in our experimental conditions.
 \subsection{Crystallographic defects induced by N implantation}
\begin{figure}
\includegraphics[width=7cm]{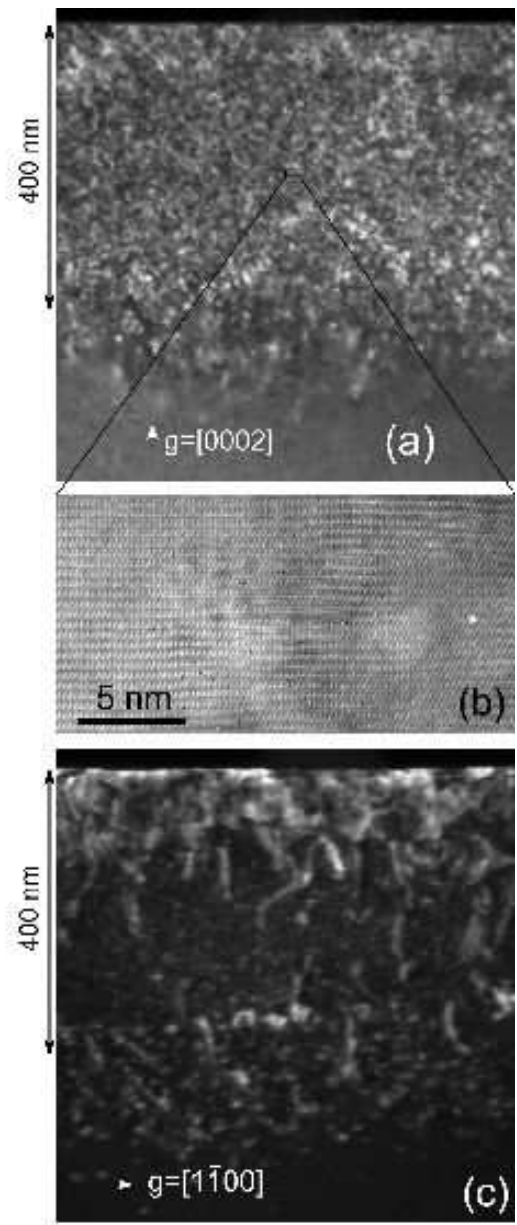}
\caption{\label{Nnr_CS_WB_HR} TEM weak-beam images of the as-implanted sample observed in cross-section with (a) $\mathbf{g}=[0002]$, and (c) $\mathbf{g}=[1\overline{1}00]$ diffracting conditions. (b) High resolution TEM image taken along the $\langle11\overline{2}0\rangle$ zone axis in the area indicated in (a).}
\end{figure}
\begin{figure}
\includegraphics[width=8cm]{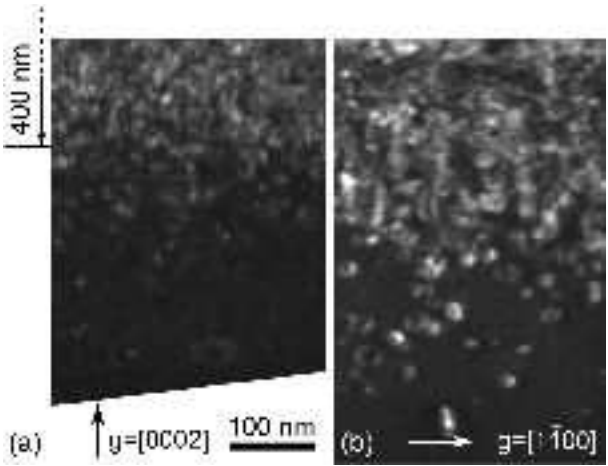}
\caption{\label{Nnr_CS_WB_zoom}Cross sectional weak-beam images of the implanted zone at 400 nm under the surface with (a) $\mathbf{g}=[0002]$, and (b) $\mathbf{g}=[1\overline{1}00]$ diffracting conditions.}
\end{figure}
Fig.~\ref{Nnr_CS_WB_HR} shows cross-sectional TEM images of defects in the as-implanted layer. Three types of defects can be identified: dislocation loops lying in basal planes; dislocation loops lying in prismatic planes, and linear dislocations. 
\subsubsection{Basal loops}
On the [0002] weak-beam image [Fig.~\ref{Nnr_CS_WB_HR}(a)], overlapping black and white contrasts due to a high density of basal loops are visible at depths between 0 and 400~nm. These depths correspond to the zone of high N concentration [Fig.~\ref{N_SIMS_SRIM}]. The habit plane of the loops in this zone was identified on cross sectional high resolution images taken along a $\langle11\overline{2}0\rangle$ zone axis, as shown in Fig.~\ref{Nnr_CS_WB_HR}(b). An additional (0001) insertion plane is detectable, indicating that the loops lie in basal planes and are of interstitial type. The mean loop diameter was determined to be 5 nm from various high resolution images. Due to the high density and the small size of these basal loops, their Burgers vector is difficult to determine without ambiguity from a $\mathbf{g}.\mathbf{b}$ analysis or from high resolution images. On the $[1\overline{1}00]$ image [Fig.~\ref{Nnr_CS_WB_HR}(c)], the basal loops show a faint contrast. It may be a residual contrast ($\mathbf{g} \cdot (\mathbf{b} \times \mathbf{u}) \neq 0$ even if $\mathbf{g}.\mathbf{b} = 0$, where $\mathbf{u}$ is a unit vector along the dislocation line), or it may correspond to a situation where $\mathbf{g}.\mathbf{b} \neq 0$. This is compatible with a Burgers vector $\mathbf{b}$ equal either to $\mathbf{c}/2$ or to $\mathbf{c}/2+\mathbf{p}$, which are the two possibilities for interstitial loops lying in basal planes in the hexagonal structure.
\subsubsection{Prismatic loops}
Some dislocation loops are visible on the $[1\overline{1}00]$ weak-beam image at depths between 400~nm and 550~nm [Fig.~\ref{Nnr_CS_WB_HR}(c)], in a region where the nitrogen concentration is lower according to SIMS measurements [Fig.~\ref{N_SIMS_SRIM}]. Fig.~\ref{Nnr_CS_WB_zoom} gives a more precise image of this zone at a depth of 400~nm. Prismatic loops are visible on $[10\overline{1}0]$ images, but not on $[0002]$ images, indicating that their Burgers vectors lie in the basal plane. Because no stacking faults are visible inside the loops on $[1\overline{1}00]$ images [Fig.~\ref{Nnr_CS_WB_HR}(c) and Fig.~\ref{Nnr_CS_WB_zoom}(b)], $\mathbf{b}$ does not have any component along $\mathbf{p}$. The Burgers vector of these loops is therefore $\mathbf{b}=\mathbf{a}$. In the literature, prismatic loops with $\mathbf{b}=\mathbf{a}$ are commonly observed in several hexagonal materials, such as quenched Mg,\cite{hillairet_am_1970} or in materials irradiated by electrons, neutrons or ions such as Zr,\cite{jostsons_jnm_1977} Co,\cite{foll_pssa_1977} SiC,\cite{wong-leung_nimprb_2008} and ZnO.\cite{yoshiie_pma_1980}
\subsubsection{Linear dislocations}
\begin{figure}
\includegraphics[width=8cm]{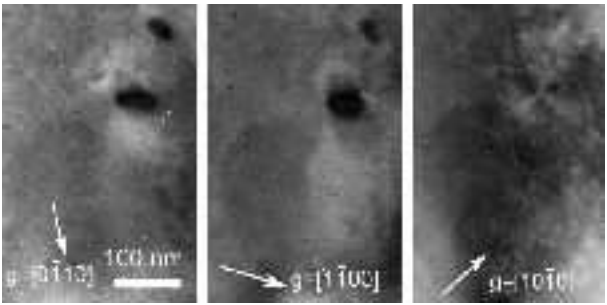}
\caption{\label{Nnr_PV_2B}Bright-field TEM images in plan-view of the implanted ZnO layer before annealing.}
\end{figure}
Some linear dislocations are visible on $[1\overline{1}00]$ images [Fig.~\ref{Nnr_CS_WB_HR}(c)], but are out of contrast on $\mathbf{g}=[0002]$ images [Fig.~\ref{Nnr_CS_WB_HR}(a)]. Consequently their Burgers vectors lie in the basal plane. They are also visible on plane-view images taken near the $\langle0001\rangle$ zone axis [Fig.~\ref{Nnr_PV_2B}]. Two linear dislocations are visible for $\mathbf{g}=[0\overline{1}10]$ and $\mathbf{g}=[1\overline{1}00]$, and are out of contrast for $\mathbf{g}=[10\overline{1}0]$. It results from this analysis that the Burgers vector of these linear dislocations is $\mathbf{b}=1/3[\overline{1}2\overline{1}0]=\mathbf{a}$. The mechanism for the formation of these linear dislocations is not clearly understood yet.
\subsection{Evolution of the defects upon annealing}
\begin{figure}
\includegraphics[width=7cm]{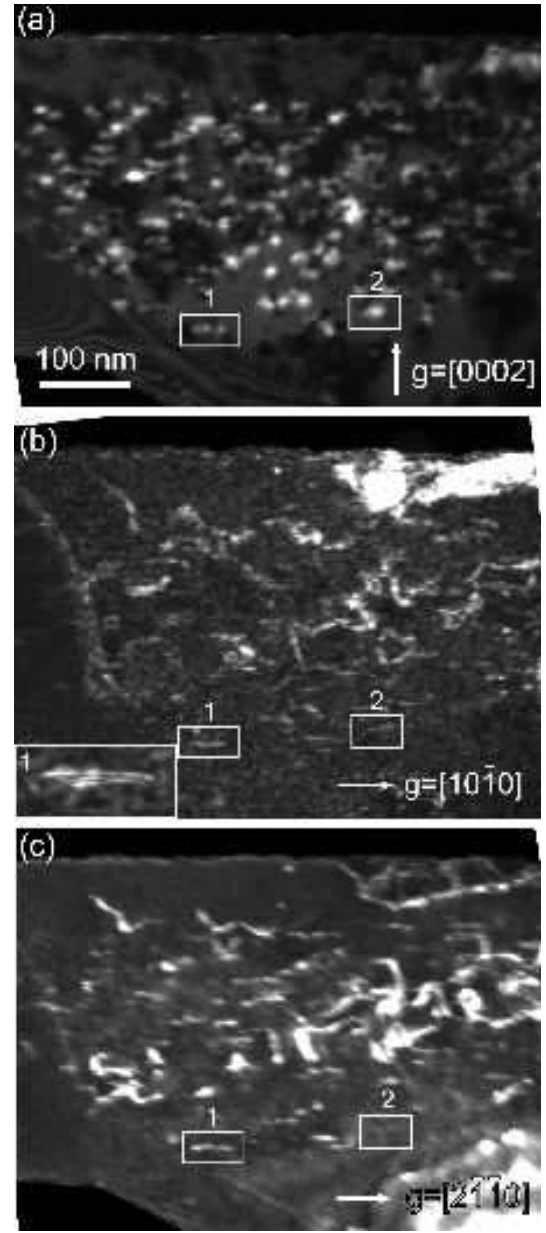}
\caption{\label{Nr_CS_WB}Weak-beam images of N implanted ZnO after annealing with (a) $\mathbf{g}=[0002]$, (b) $\mathbf{g}=[10\overline{1}0]$, and (c) $\mathbf{g}=[2\overline{1}\overline{1}0]$ diffracting conditions.}
\end{figure}
\begin{figure}
\includegraphics[width=7cm]{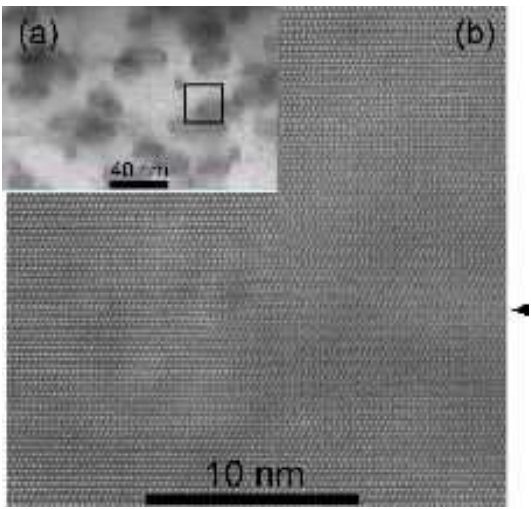}
\caption{\label{Nr_CS_HR}(a) Low magnification bright-field scanning TEM image of basal loops after annealing, and (b) high-resolution scanning TEM image of the side of one loop (corresponding to the area marked by a rectangle in (a)) taken along a $\langle11\overline{2}0\rangle$ zone axis with a high-angle annular dark-field detector.}
\end{figure}
After annealing at $900\,^{\circ}\text{C}$ during 15~minutes, basal loops and linear dislocations are still observed, but prismatic loops are no longer visible [Fig.~\ref{Nr_CS_WB}]. Basal loops have grown in diameter and their density has decreased. Their mean diameter has increased from 5~nm to 21~nm. Another noticeable effect related to annealing is the complete disappearance of loops in the subsurface region over a depth of 65~nm.
The decreasing density of basal loops together with their larger size allow us to determine their Burgers vectors more precisely than before annealing. They can be determined on cross-sectional weak-beam images observed with three different diffracting conditions: $\mathbf{g}=[0002]$, $\mathbf{g}=[10\overline{1}0]$ and $\mathbf{g}=[2\overline{1}\overline{1}0]$ [Fig.~\ref{Nr_CS_WB}]. All basal loops are visible for $\mathbf{g}=[0002]$ and $\mathbf{g}=[10\overline{1}0]$. Thus $\mathbf{b}$ has components parallel and perpendicular to the $\mathbf{c}$ axis. The same number of loops is visible for $\mathbf{g}=[10\overline{1}0]$ and $\mathbf{g}=[0002]$, while only about two-thirds of the basal loops are visible for $\mathbf{g}=[2\overline{1}\overline{1}0]$. For example, the loop marked 1 in Fig.~\ref{Nr_CS_WB} is visible with the three different diffracting conditions whereas the loop marked 2 is visible for $\mathbf{g}=[10\overline{1}0]$ and $\mathbf{g}=[0002]$ but not for $\mathbf{g}=[11\overline{2}0]$. It can be concluded that the Burgers vector of basal loops after annealing is $\mathbf{b} = \mathbf{c}/2 + \mathbf{p}$. Moreover Moiré fringes inside the loops on the $[10\overline{1}0]$ image indicate the presence of a stacking fault with a displacement vector along $\mathbf{p}$. A zoom on one basal loop with such fringes is shown in the inset of Fig.~~\ref{Nr_CS_WB}(b). A low magnification bright-field scanning TEM image of the basal loops is shown on Fig.~\ref{Nr_CS_HR} (a). Fig.~\ref{Nr_CS_HR} (b) is a zoom on the edge of the basal loop marked in Fig.~\ref{Nr_CS_HR} (a). One extra atomic plane inside the dislocation loop is visible on this high resolution STEM image: the basal loops are of interstitial type as before annealing.
Some linear dislocations crossing the implanted layer are still visible on $[2\overline{1}\overline{1}0]$ images, but some of them are out of contrast for $\mathbf{g} = [10\overline{1}0]$, and none are visible on $\mathbf{g}=[0002]$ images. Consequently they still have a $\mathbf{b} = \mathbf{a}$ Burgers vector. 
\subsection{Deformations in the implanted zone as measured by XRD}
\label{XRD_results}
\begin{figure}
\includegraphics[width=8cm]{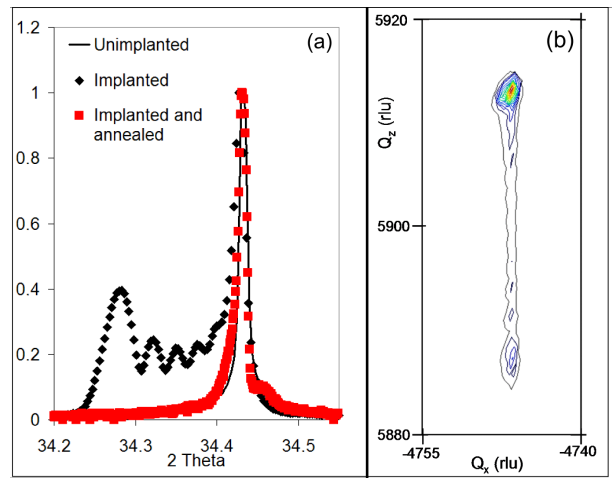}
\caption{\label{N_XRD_exp} XRD results. (a) $\theta / 2 \theta$ scans on the symmetrical [0002] reflection obtained on ZnO substrates before implantation, after N implantation, and after implantation and annealing. (b) Reciprocal space map on a $[11\overline{2}4]$ asymmetric peak obtained on the as-implanted sample}
\end{figure}
Fig.~\ref{N_XRD_exp} shows the results obtained by high resolution XRD. Fig.~\ref{N_XRD_exp} (a) displays $\theta/2\theta$ scans done on the symmetrical [0002] reflection for an unimplanted sample, the as-implanted sample, and the implanted and annealed sample. Fringes on the left of the [0002] peak indicates that nitrogen implantation in ZnO induces a deformed layer with a larger \textit{c} parameter. Reciprocal space maps on the asymmetrical $[11\overline{2}4]$ reflection [Fig.~\ref{N_XRD_exp} (b)] demonstrate that the dilatation of the implanted layer occurs only in the c direction, with no in-plane deformation. This gives evidence of an increase of the ZnO unit cell induced by implantation. Goland and Keating have already demonstrated that interstitial dislocation loops in hexagonal structures yield an increase of the lattice parameter in the direction perpendicular to the habit plane of the loops.\cite{goland_jap_1970} This phenomenon, intensively studied in nuclear material science, was called "irradiation growth".\cite{bullough_jnm_1980} In our study, we infer that the basal loops yield an increase of the c parameter. The deformation induced by the prismatic loops is not detectable, as deduced from Fig.~\ref{N_XRD_exp} (b).
\begin{figure}
\includegraphics[width=8cm]{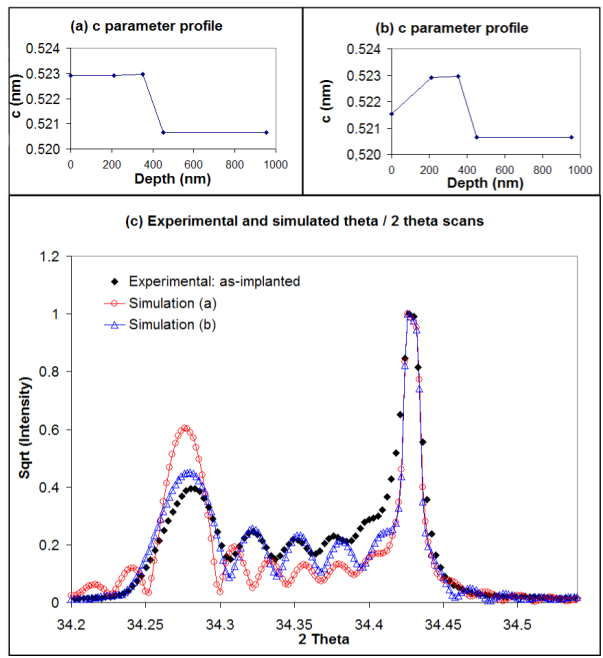}
\caption{\label{N_XRD_simul} (c) $\theta / 2 \theta$ scan on the symmetrical [0002] reflection obtained on the as-implanted ZnO substrate, and XRD simulations for various d-spacing gradients along the c axis: (a) uniform deformation from 0 to 400~nm, and (b) deformation profile proportional to the implantation profile (see Fig.~\ref{N_SIMS_SRIM}).}
\end{figure}
Simulations of rocking curves based on kinematical diffraction are shown in Fig.~\ref{N_XRD_simul}(c). The experimental rocking curve is initially fitted with a uniform deformation profile as represented in the inset of Fig.~\ref{N_XRD_simul}(a). This corresponds to a layer of 400 nm with a 0.4\% uniaxial deformation. However, a better fit of the experimental rocking curve is obtained with a deformation profile which follows qualitatively the implantation profile, as represented in the inset of Fig.~\ref{N_XRD_simul}(b). This suggests that the distribution of the basal loops follows roughly the implantation profile.
The XRD scan for the annealed sample is also presented in Fig.~\ref{N_XRD_exp}(a): a small deformation of only 0.05\% remains after annealing, which can be ascribed to the remaining basal loops.
\section{Discussion}
\begin{table*}
\caption{\label{table1}Types of dislocation loops observed by TEM in ion or electron irradiated ZnO before and after annealing. The N concentration is noted $C_{N}$.}
\begin{ruledtabular}
\begin{tabular}{p{2cm}|p{4.5cm} p{4.5cm} p{4.5cm}}
& \textbf{Yoshiie \textit{et al.}}\cite{yoshiie_pma_1980, yoshiie_pma_1983} 
& \textbf{Couderc \textit{et al.}}\cite{couderc_mmm_1995}
& \textbf{This study} 
\\
& Electron irradiation
& Ar implantation
& N implantation
\\
&700~keV
&5~keV
&50~keV to 200~keV
\\
&Pure ZnO
&Zn\textsubscript{97.4}Bi\textsubscript{1}Mn\textsubscript{1}Ti\textsubscript{0.6}O
&ZnO with $ 0 < C_{N} < 0.22 \% $
\\\hline
Implanted sample
& Interstitial basal loops with
& Interstitial basal loops with
& Interstitial basal loops with
\\
&$\mathbf{b}=\mathbf{c}/2$
&$\mathbf{b}=\mathbf{c}/2 + \mathbf{p}$
&$\mathbf{b}=\mathbf{c}/2$ or $\mathbf{b}=\mathbf{c}/2 + \mathbf{p}$ (ambiguity) for high $C_{N}$ region
\\
& Prismatic loops with $\mathbf{b}=\mathbf{a}$
& 
& Prismatic loops with $\mathbf{b}=\mathbf{a}$ for low $C_{N}$ region
\\\hline
Implanted and annealed sample
& Fewer prismatic loops. No more loops if $T > 800\,^{\circ}\text{C}$.
& 
& Interstitial basal loops with $\mathbf{b}=\mathbf{c}/2+\mathbf{p}$. Prismatic loops and loops close to the surface  are removed at $900\,^{\circ}\text{C}$. 
\end{tabular}
\end{ruledtabular}
\end{table*}
Table~\ref{table1} sums up the TEM observations of implantation defects obtained by Yoshiie \textit{et al.}\cite{yoshiie_pma_1980,yoshiie_pma_1983} for electron irradiation of pure ZnO thin foils, by Courderc \textit{et al.}\cite{couderc_mmm_1995} for Ar implantation of Bi-Mn-Ti-doped ZnO, and by ourselves for N implantation. All the observed basal dislocation loops are of interstitial type, whereas prismatic loops are only assumed to be interstitial.
\subsection{Formation of dislocations during implantation}
\begin{table}
\caption{\label{table2}Migration barriers and temperature at which ZnO native defects become mobile (from Janotti \textit{et al.}\cite{janotti_prb_2007}).}
\begin{ruledtabular}
\begin{tabular}{l l}
Zinc interstitial		& 0.57~eV ($-54\,^{\circ}\text{C}$) \\
Oxygen interstitial	& 0.87~eV ($62\,^{\circ}\text{C}$) or 1.14~eV ($166\,^{\circ}\text{C}$) \\
Zinc vacancy				& 1.40~eV ($266\,^{\circ}\text{C}$) \\
Oxygen vacancy			& 2.36~eV ($636\,^{\circ}\text{C}$) \\
\end{tabular}
\end{ruledtabular}
\end{table}
\subsubsection{Formation of interstitial dislocation loops}
First-principles calculations show that native point defects are very mobile in ZnO.\cite{janotti_prb_2007, erhart_apl_2006, erhart_prb_2006} Calculated migration barriers and the corresponding temperatures at which native defects become mobile (considering that a point defect is mobile if its jump rate is more than 1 per second) are reported in Table~\ref{table2}.\cite{janotti_prb_2007} It appears that at room temperature, interstitials are very mobile, and vacancies almost immobile. In ion implanted ZnO, Frenkel pairs are formed by a cascade damage. Some of them annihilate, but mobile interstitials can agglomerate to form dislocation loops. Immobile vacancies stay isolated or may form vacancy clusters of a few vacancies.\cite{chen_jpcm_2004,borseth_prb_2008,chen_apl_2005,dong_prb_2010} Our observations of interstitial loops are thus coherent with theoretical predictions. Concerning previous TEM investigations (reported in Table~\ref{table1}), Yoshiie \textit{et al.} and Couderc \textit{et al.} also observed that basal loops were of interstitial type.
\subsubsection{Type of dislocation loops according to N concentration}
Let us discuss about the parameters that may determine the habit planes of interstitial loops. Föll \textit{et al.} proposed a rule to predict the habit planes of dislocation loops in hexagonal closed-packed structures depending on the $c/a$ ratio.\cite{foll_pssa_1977} If $d_{\{0001\}} / d_{\{10\overline{1}0\}} < 1$ (where d is the lattice plane spacing), which is equivalent to $c/a < \sqrt{3} = 1.732$, the loops should form in the $\{10\overline{1}0\}$ prismatic planes. If $c/a > 1.732$, the loops should form in basal planes. Griffiths discussed more precisely the habit planes of interstitial loops in a review on the microstructure of irradiated hexagonal compact metals.\cite{griffiths_jnm_1993} Föll's rule is respected for pure materials. But for impure materials with $c/a < 1.732$, both prismatic and basal loops are observed. In particular, dislocation loops formed by quenching in Mg (for which $c/a = 1.623 $) were studied.\cite{hillairet_am_1970} Prismatic loops where observed for Mg with 0.0001\% impurities, while basal loops were observed for Mg with 0.05\% impurities. In the ZnO case, prismatic loops should form preferentially, as $c/a = 1.602 $. However, taking into account Griffiths' observations, the number of basal loops should increase relatively to the number of prismatic loops when the impurity concentration increases. This is in agreement with the results for irradiated ZnO summarized in Table~\ref{table1}: prismatic loops together with basal loops are present in relatively pure material (Yoshiie \textit{et al.}) whereas only basal loops are present in impure materials (Couderc \textit{et al.}). In our study, basal loops are observed in the high N concentration region and prismatic loops in the low N concentration region, in agreement with Griffiths' observations.
This dependence of the habit plane with the impurity concentration can be explained by considering the energy of a dislocation loop. The total elastic energy of a loop is $E=E_l \times l + \gamma \times S$ where $E_l$ is the energy per unit-length of the dislocation line, $l$ the length of the dislocation, $\gamma$ the energy of the stacking fault, and $S$ its surface. A general result for a large variety of materials is that the stacking fault energy is lowered when the impurity concentration is increased.\cite{gallagher_mt_1970} For example, in the case of wurtzite materials like GaN, the density of basal stacking faults is increased by a factor of 10 with Si doping.\cite{molina_apl_1999} \textit{Ab-initio} calculations indicate this is due to a lower basal stacking fault energy resulting from the Si doping.\cite{chisholm_jcg_2001} Because prismatic loops have a Burgers vector $\mathbf{b} = \mathbf{a}$, and thus no stacking fault inside (the dislocation is perfect), their energy does not change when the impurity concentration increases. But the basal loops with $\mathbf{b} = \mathbf{c}/2 + \mathbf{p}$ or $\mathbf{b} = \mathbf{c}/2$ contain a stacking fault: their energy is lowered when the impurity concentration is increased. Thus basal loops become more stable than prismatic loops for high impurity concentration.
The Burgers vectors of basal loops may also be accounted for by the impurity concentration. In our study, and in Couderc \textit{et al.}'s one with impure ZnO, basal loops have $\mathbf{b} = \mathbf{c}/2 + \mathbf{p}$ with a type one intrinsic stacking fault. In Yoshiie \textit{et al.}'s study with pure ZnO, the Burgers vector is $\mathbf{b} = \mathbf{c}/2$ with an extrinsic stacking fault. To explain these experimental observations, we propose that a high impurity concentration may lower the intrinsic stacking fault energy more than the extrinsic one and thus favours $\mathbf{b} = \mathbf{c}/2 + \mathbf{p}$ in impure materials.
\subsection{Recovery of implanted ZnO with annealing}
\subsubsection{Behaviour of dislocation loops}
As observed for many crystals (for example Si),\cite{bonafos_jap_1998} and for ZnO in particular,\cite{yoshiie_pma_1980} loop diameters increase upon annealing while their density decreases. As two dislocation loops attract each other, loops tend to coalesce.\cite{hirth_1992} Because the Burgers vector of the loops do not lie in the loop plane, this coalescence does not occur by glide. It occurs rather by climb, which is a diffusion-assisted phenomenon.
The disappearance of all loops in the sub-surface region (up to 65~nm under the surface) with annealing is observed. One could attribute this to the weaker density of dislocation loops in this region, as deduced from the XRD study (see \ref{XRD_results}). But it is not sufficient to explain the complete disappearance of the loops. The specific role of surface must be considered. Actually, it was already observed that the density of loops produced by implantation in silicon decreases when the implantation depth approaches the surface.\cite{raman_apl_1999} This was ascribed to the diffusion of interstitials towards the surface. Sub-surface area free of dislocation loops were also observed in hexagonal materials such as Zr alloy,\cite{woo_pma_1991} Mg,\cite{hillairet_am_1970} and CdTe.\cite{leo_mseb_1993} Because hexagonal materials are anisotropic, it was pointed out that all surfaces do not play the same role.\cite{woo_pma_1991} In ZnO, irradiated thin foils\cite{yoshiie_pma_1980} and Ga-implanted ZnO nanorods\cite{yao_jap_2009} become free of dislocations after annealing at $800\,^{\circ}\text{C}$. These examples stress the role of surfaces to remove dislocation loops. Surfaces act as a sink for point defects and thus causes the "dissolution" of dislocation loops.
In this study, the disappearance of prismatic loops is observed for an annealing at $900\,^{\circ}\text{C}$. Yoshiie \textit{et al.} observed that the basal loops to prismatic loops ratio increased with the irradiation temperature from $200\,^{\circ}\text{C}$ to $500\,^{\circ}\text{C}$. Extrapolation of their results shows that no prismatic loops should be present after $720\,^{\circ}\text{C}$, which is coherent with our study. Preferential nucleation of basal loops was also observed in Zr alloys at higher temperature.\cite{griffiths_jnm_1993,jostsons_jnm_1977} A lower basal stacking fault energy due to a higher temperature\cite{gallagher_mt_1970} may cause prismatic loops to be unstable and disappear.
\subsubsection{Considerations to improve structural recovery}
After annealing, the XRD analysis indicates a good structural recovery with a residual deformation of only 0.05\%. However, PL spectra show no donor-acceptor pair emission when carried out directly on the sample surface. But when the top-surface layer (those without defects on a depth of 65 nm) is removed by chemical-mechanical polishing, donor-acceptor emission is evidenced by PL.\cite{marotel_pc_2010} Hall measurement using the Van der Paw configuration was realized on the annealed sample after removal of the surface layer. However, the implanted layer was found too resistive to be measured, most probably because of the remaining dislocations observed in TEM. In our view, the electroluminescence observed in the literature for N-implanted ZnO annealed at $800\,^{\circ}\text{C}$ is possibly due to a metal-insulator-semiconductor junction.\cite{adekore_jap_2007} To conclude, a better structural recovery is still needed to produce light-emitting devices with N implanted ZnO as a \textit{p}-type layer.
Several solutions may be envisaged for a better structural recovery of the implantation induced defects. \begin{itemize}
\item A longer annealing, or an annealing at higher temperature are possible, but it would lead to the disappearance of donor-acceptor pair emission,\cite{marotel_pc_2010} and to the possible deactivation of N doping by forming N\textsubscript{2} molecules.\cite{fons_prl_2006}
\item Lower implantation doses would produce less defects, and consequently the annealed material would have less defects too. Moreover, more prismatic loops should form for low N dose, and they should disappear with annealing. But the dose must remain sufficient to incorporate enough active dopants.
\item Finally, implantation could be carried out on ZnO nanowires to take advantage of a better recovery due to the larger surface to volume ratio in these structures. Actually it was observed that structural defects were removed in Ga implanted nanowires after an annealing at $800\,^{\circ}\text{C}$.\cite{yao_jap_2009} Moreover, UV electroluminescence was obtained from a p-n junction in ZnO nanowires, with arsenic implantation doping.\cite{yang_apl_2008} \end{itemize}
\section{Summary}
N implantation in ZnO at room temperature produces immobile vacancies and mobile interstitials. Interstitials agglomerate into dislocation loops lying in basal and prismatic planes. For depths between 0 and 400 nm, where N concentrations are high, loops lying in basal planes are observed. Deeper than 400 nm, where N concentration decreases, loops lying in prismatic planes with $\mathbf{b} = \mathbf{a}$ are observed. Linear dislocations with $\mathbf{b} = \mathbf{a} $ are also observed. The implanted layer is shown by XRD to have a larger c parameter, which is ascribed to the predominant basal loops. After annealing at $900\,^{\circ}\text{C}$, basal loops with $\mathbf{b}= \mathbf{c}/2 + \mathbf{p}$ and linear dislocations with $\mathbf{b} = \mathbf{a} $ are still observed, but prismatic loops disappear. A mechanism is proposed to explain these phenomena: basal loops are favoured against prismatic loops at high N concentration or high temperature because the basal loop stacking fault energy decreases. With annealing, basal loops coalesce: they are less numerous and bigger. They disappear below the surface. Both phenomena are ascribed to diffusion of point defects. Although XRD and PL results seems to show a good recovery of the material, the remaining basal loops observed in TEM lead to a non conductive layer, as demonstrated by Hall effect measurements. Propositions are made to improve the structural recovery of N-implanted ZnO.
\begin{acknowledgments}
The authors would like to acknowledge Carole Granier for her precious technical assistance, Frédéric Milesi for implantations, Jean-Paul Barnes for making some corrections to the English of this publication, and the french national research agency (ANR) for funding through the Carnot program (2006/2009).
\end{acknowledgments}
\bibliography{implantation_ref}
\end{document}